\documentclass{article}%
\usepackage{amsmath}
\usepackage{amsfonts}
\usepackage{amssymb}
\usepackage{graphicx}%
\newcommand{\bpartial}{\mathop{\partial\kern -4pt\raisebox{.8pt}{$|$}}}
\newcommand{\bra}{\mathopen{[\kern-1.6pt[}}
\newcommand{\ket}{\mathclose{]\kern-1.5pt]}}
\newcommand{\bbra}{\mathopen{[\kern-2.2pt[\kern-2.3pt[}}
\newcommand{\bket}{\mathclose{]\kern-2.1pt]\kern-2.3pt]}}

\makeindex
\begin{document}
\title{\bf Complex and biHermitian structures on four dimensional real Lie algebras }
\author { A. Rezaei-Aghdam \hspace{-3mm}{ \footnote{e-mail: rezaei-a@azaruniv.edu}\hspace{2mm} {\small and }
M. Sephid \hspace{-3mm}{ \footnote{ e-mail: s.sephid@azaruniv.edu}}\hspace{2mm}} \\
{\small{\em Department of Physics, Faculty of science, Azarbaijan University }}\\
{\small{\em of Tarbiat Moallem , 53714-161, Tabriz, Iran  }}}

\maketitle
\begin{abstract}
We give a new method for calculation of complex and biHermitian
structures on low dimensional real Lie algebras. In this method,
using non-coordinate basis, we first transform the Nijenhuis
tensor field and biHermitian structure relations on Lie groups to
the tensor relations on their Lie algebras. Then we use adjoint
representation for writing these relations in the matrix form; in
this manner by solving these matrix relations and using
automorphism groups of four dimensional real Lie algebras we
obtain and classify all complex and biHermitian structures on four
dimensional real Lie algebras.
\end{abstract}
\newpage
\section{\bf Introduction}
Calculation of complex structures on homogeneous complex
manifolds and especially on Lie groups is important from both
mathematical and physical point of view. Mathematically,
classification of these manifolds is based on determination of
possible complex structures. From the physical point of view
these structures have an important role in the N=(2,2)
supersymmetric sigma models \cite{G}. It is shown that the N=(2,2)
extended supersymmetry in sigma model implies the existence of
biHermitian structure on the target manifold such that the
complex structures are covariantly constant with respect to
torsionful affine connections (see for example \cite{SL} and
references therein). Furthermore, it is shown that the algebraic
structures related to these biHermitian structures for N=(2,2)
supersymmetric WZW models are the Manin triples
\cite{Sp},\cite{L}. For these reasons, the calculation of the
complex and the biHermitian structures on manifolds especially on
Lie groups, are important. Samelson \cite{Sa} shows that compact
Lie groups always admit an invariant complex structure . As for
the non-compact case, Morimoto \cite{Mo} proves that there always
exist invariant complex structures on any even dimensional
reductive Lie groups. In \cite{S} and \cite{O}, complex structures
on real four dimensional Lie algebras are classified. The method
used in this works is special and seems not to be adequate for the
calculation in higher dimensions. In the present paper we give a
new method for this purpose, which can be applied for low
dimensional Lie groups . In this method, using non-coordinate
basis, we first transform the Nijenhuis tensor on Lie groups to
algebraic tensor relations on their Lie algebras. Then, using
adjoint representation we rewrite these relations in the matrix
form. Finally, we solve these matrix relations using Maple. In
this research we perform this for real four dimensional Lie
algebras. The results for some algebras are different and complete
with respect to \cite{O}. Furthermore, calculation of biHermitian
structure for four dimensional Lie algebras is new .\\
The paper is organized as follows. In section 2, using
non-coordinate basis, we transform the Nijenhuis tensor relation
on a Lie group to the algebraic tensor relation on its Lie
algebra. Then, using adjoint representation, we write these
relations in the matrix form. The relations can also be obtained
from definition of complex structures on Lie algebras as
endomorphism of them. Then, in section 3 using Maple we solve
these matrix relations to obtain complex structures on real four
dimensional Lie algebras. In this process, we apply automorphism
groups of real four dimensional Lie algebras for obtaining
non-equivalent complex structures (table 1). We then compare our
results with \cite{S} and \cite{O}. Note that here we use Patera
{\it etal} \cite{P} classification of real four dimensional Lie
algebras. The list of Lie algebras and their automorphism groups
\cite{PaP} are given in appendix. In section 4, we first
transform the tensorial form of the biHermitian relations on Lie
groups into the algebraic tensorial relations on their Lie
algebras. In this respect, we define biHermitian structure on Lie
algebra independently and give an equivalent relation for
obtaining non equivalent biHermitian structures. Then using
adjoint representation we rewrite these relations in the matrix
forms and solve them by Maple. Therefore, the present paper will
be a continuation to the discussion of biHermitian structure on
real four dimensional Lie algebras. Some discussions are given in
the conclusion section.
\section{\bf A brief review of complex structures on Lie groups}
{\bf Definition 1:} {\it Let M be a differentiable manifold, then
the pair (M,J) is called almost complex manifold if there exists a
tensor field J of type (1,1) such that at each point p of M ,
$J_{p}^2=-1$; tensor field J is also called the almost complex
structure. Furthermore, if the Lie bracket of any vector fields
of type $(1,0)$ $X,Y\in T_{p}{M^{+}}$ is again of the same type,
then the complex structure $J_{p}$ is said to be integrable, where
$T_{p}{M^{+}} = \{Z \in T_{p}{M^{C}} \mid J_{p}Z=+iZ \}$.}\\

\hspace{-.7cm} {\bf Theorem (Newlander and Nirenberg \cite{NN}) :}
 {\it An almost complex structure J on a manifold M is integrable
if and only if
\begin{equation}
 N(X,Y)=0 \hspace{1cm}\forall X,Y \in \chi(M)\hspace{1mm},
\end{equation}
 where ${\chi(M)}$ is the set of vector fields on M and the
Nijenhuis tensor $N: \chi(M) \otimes \chi(M) \rightarrow \chi(M)$
is given by
\begin{equation}
N(X,Y)=[X,Y]+J[JX,Y]+J[X,JY]-[JX,JY]\hspace{1mm}.
\end{equation}}
In the coordinate basis, i.e the basis
$\{{e_{\mu}=\frac{\partial}{\partial{x}^{\mu}}}\}$ and
$\{{dx^{\mu}}\}$ for vectors and dual vectors (forms) respectively
on M, the almost complex structures and Nijenhuis tensor are
expressed as $J=J_{\mu}\hspace{0cm}^{\nu} e_{\nu}\otimes
dx^{\mu}$  and \\ $N=N_{\mu\nu}\hspace{0cm}^{\lambda}
dx^{\mu}\otimes dx^{\nu} \otimes e_{\lambda}$ respectively and
the integrability condition $(2)$ can be rewritten as follows :
\begin{equation}
N_{\kappa\nu}\hspace{0cm}^{\mu}
=J_{\lambda}\hspace{0cm}^{\mu}(\partial_{\kappa}J_{\nu}\hspace{0cm}^{\lambda})+J_{\nu}\hspace{0cm}^{\lambda}(\partial_{\lambda}J_{\kappa}\hspace{0cm}^{\mu})
-J_{\lambda}\hspace{0cm}^{\mu}(\partial_{\nu}J_{\kappa}\hspace{0cm}^{\lambda})-J_{\kappa}\hspace{0cm}^{\lambda}(\partial_{\lambda}J_{\nu}\hspace{0cm}^{\mu}\hspace{1mm}=0.
\end{equation}
Meanwhile, the relation $J^2=-1$ can be rewritten as
\begin{equation}
J_{\lambda}\hspace{0cm}^{\mu}J_{\mu}\hspace{0cm}^{\nu}=-\delta_{\lambda}\hspace{0cm}^{\nu}.
\end{equation}
Furthermore one can rewrite  the above equations using the
non-coordinate basis $\{{\hat{e}_{\alpha}}\}$ and
$\{{\hat{\theta}^{\alpha}}\}$ for vectors and forms on M. For
these basis we have
\begin{equation}
\hat{e}_{\alpha}=e_{\alpha}\hspace{0cm}^{\mu}e_{\mu}\hspace{1cm},
\hspace{1cm}e_{\alpha}\hspace{0cm}^{\mu}\in GL(m,R)\hspace{1mm},
\end{equation}
where for the vierbeins $e_{\alpha}\hspace{0cm}^{\mu}$ and its
inverse $e^{\alpha}\hspace{0cm}_{\mu}$ we have
\begin{equation}
e^{\alpha}\hspace{0cm}_{\mu}e_{\beta}\hspace{0cm}^{\mu}=\delta^{\alpha}\hspace{0cm}_{\beta}\hspace{2mm},\hspace{2mm}
e^{\alpha}\hspace{0cm}_{\mu}e_{\alpha}\hspace{0cm}^{\nu}=\delta_{\mu}\hspace{0cm}^{\nu}\hspace{1mm}.
\end{equation}
The dual basis $\{{\hat{\theta^{\alpha}}}\}$ are defined by
$<\hat{\theta^{\alpha}},\hat{e}_{\beta}>=\delta^{\alpha}\hspace{0cm}_{\beta}$
and we have
$\hat{\theta^{\alpha}}=e^{\alpha}\hspace{0cm}_{\mu}dx^{\mu}$\hspace{1mm}.
Furthermore, the vierbeins satisfy the following relation:
\begin{equation}
{f_{\alpha\beta}}^{\gamma}=e^{\gamma}\hspace{0cm}_{\nu}(e_{\alpha}\hspace{0cm}^{\mu}
\partial_{\mu}e_{\beta}\hspace{0cm}^{\nu}-e_{\beta}\hspace{0cm}^{\mu}\partial_{\mu}
e_{\alpha}\hspace{0cm}^{\nu})\hspace{1mm},
\end{equation}
if M is a Lie group manifold G then
$f_{\alpha\beta}\hspace{0cm}^{\gamma}$,s are structure constants
of Lie algebra {\bf g} of G. Now on these bases the tensor
$J=J_{\mu}\hspace{0cm}^{\nu} e_{\nu}\otimes dx^{\mu}$ can be
rewritten as
\begin{equation}
J_{\mu}\hspace{0cm}^{\nu}
=e^{\alpha}\hspace{0cm}_{\mu}J_{\alpha}\hspace{0cm}^{\beta}e_{\beta}\hspace{0cm}^{\nu}\hspace{1mm},
\end{equation}
where $J_{\alpha}\hspace{0cm}^{\beta}$ is an endomorphism of {\bf
g}, i.e. $J:{\bf g}\longrightarrow {\bf g}$\hspace{1mm}.
 Now by applying this relation to $(4)$ we have the following matrix
 relation for matrices $J_{\alpha}\hspace{0cm}^{\beta}$:
\begin{equation}
J^{2}=-I\hspace{1mm}.
\end{equation}
Furthermore, by applying relations $(7)$ and $(8)$ to tensor
equation $(3)$ and using $(6)$ and assuming that
$J_{\alpha}\hspace{0cm}^{\beta}$ and $g_{\alpha\beta}$ are
independent of coordinates of G, after some calculations we have
the following algebraic relation for$(3)$:
\begin{equation}
f_{\beta
\alpha}\hspace{0cm}^{\gamma}+J_{\sigma}\hspace{0cm}^{\gamma}\hspace{1mm}J_{\alpha}\hspace{0cm}^{\delta}\hspace{1mm}f_{\beta
\delta}\hspace{0cm}^{\sigma}-J_{\beta}\hspace{0cm}^{\sigma}\hspace{1mm}J_{\alpha}\hspace{0cm}^{\delta}\hspace{1mm}f_{\sigma
\delta}\hspace{1mm}^{\gamma}
+J_{\beta}\hspace{0cm}^{\sigma}\hspace{1mm}J_{\delta}\hspace{0cm}^{\gamma}\hspace{1mm}f_{\sigma
\alpha}\hspace{0cm}^{\delta}=0.
\end{equation}
Finally, using adjoint representations
\begin{equation}
f_{\beta\alpha}\hspace{0cm}^{\gamma}=-({\cal{Y}}^{\gamma})_{\beta\alpha}\hspace{4mm},\hspace{4mm}f_{\beta\alpha}\hspace{0cm}^{\gamma}=-(\chi_{\beta})_{\alpha}\hspace{0cm}^{\gamma},
\end{equation}
the relation $(10)$ will have the following matrix form:
\begin{equation}
{\cal{Y}}^{\alpha}+J\hspace{1mm}{\cal{Y}}^{\beta}\hspace{1mm}J_{\beta}\hspace{0cm}^{\alpha}+J_{\beta}\hspace{0cm}^{\alpha}\hspace{1mm}{\cal{Y}}^{\beta}\hspace{1mm}J^{t}-J\hspace{1mm}{\cal{Y}}^{\alpha}\hspace{1mm}J^{t}=0
\hspace{1mm},
\end{equation}
or
\begin{equation}
{\chi}_{\alpha}+J\hspace{1mm}{\chi}_{\alpha}\hspace{1mm}J+J_{\alpha}\hspace{0cm}^{\beta}\hspace{1mm}{\chi}_{\beta}\hspace{1mm}J-J_{\alpha}\hspace{0cm}^{\beta}\hspace{1mm}J\hspace{1mm}{\chi}_{\beta}=0
\hspace{1mm}.
\end{equation}
Note that the above equation can also be obtained from the
definition of complex structure on Lie algebra {\bf g} as follows.\\
\hspace{-8mm} {\bf Definition 2:} {\it An integrable complex
structure on a real Lie
algebra\hspace{1mm}{\bf g} is an endomorphism J of {\bf g} such that\\
\begin{equation}
\hspace{-13.8cm}a) \hspace{1mm}J^{2}=-Id ,\\
\end{equation}
\begin{equation}
\hspace{-5.5cm}b)
\hspace{1mm}[X,Y]+J[JX,Y]+J[X,JY]-[JX,JY]=0\hspace{1mm},
\hspace{1cm}\forall X,Y\in {\bf g}\hspace{1mm}.
\end{equation}}
Now if we use $\{X_{\alpha}\}$ as basis for Lie algebra {\bf g}
with the following structure constants:
\begin{equation}
[X_{\alpha},X_{\beta}]=f_{\alpha\beta}\hspace{0cm}^{\gamma}
X_{\gamma}\hspace{1mm},
\end{equation}
and use the following relation for J:
\begin{equation}
JX_{\alpha}=J_{\alpha}\hspace{0cm}^{\beta}X_{\beta}\hspace{1mm},
\end{equation}
then relations $(14)$ and $(15)$ can be rewritten as $(9),(12)$ or
$(13)$. Now, in order to obtain algebraic complex structures
$J_{\alpha}\hspace{0cm}^{\beta}$, it is enough to solve equations
$(9)$ and $(12)$ or $(13)$ simultaneously\hspace{1mm}. We do this
for real four dimensional Lie algebras in the next section.

\vspace{6mm}
\section{\bf Calculation of complex structures on four dimensional real Lie algebras}
In this section we use the Patera  {\it etal} classification
\cite{P} of four dimensional real Lie algebras . The commutation
relations and the automorphism groups of these Lie algebras are
given in the appendix{ \footnote{ Note that for decomposable Lie
algebra $(L_{3}\oplus R)$ we use Bianchi classification of real
three dimensional Lie algebras $L_{3}$.}}. Now one can write the
adjoint representation $({\cal Y})$ for these Lie algebras and
then solve the matrix relations $(9)$ and $(12)$ for obtaining
complex structures. We do this by Maple. Note that in this
process one can obtain equivalent complex structures; to avoid
these and in order to obtain inequivalent complex structure we
use the following equivalent relation:

\bigskip

{\it{\bf Definition 3\cite{O}} : Two complex structures $J_{1}$
and $J_{2}$ of Lie algebra {\bf g} are equivalent if there exists
an element A of automorphism group of Lie algebra {\bf g}
(Aut({\bf g})) such that:}
\begin{equation}
J_{2}=A\hspace{1mm} J_{1} \hspace{1mm} A^{-1}\hspace{1mm}.
\end{equation}
Note that this relation is an equivalent relation.\\
In this way, we do this and obtain all non-equivalent complex
structures on four dimensional real Lie algebras. The results are
classified in table 1 { \footnote{Note that in this table the
bases are shown with $\{e_{\alpha}\}$ instead of
$\{X_{\alpha}\}$}}. As the table shows 21 out of 30 real four
dimensional Lie algebras have complex structures. To compare
these results with results of \cite{O} first we must obtain the
isomorphism relations between the four dimensional real Lie
algebras presented in \cite{P} and those presented in \cite{O}.
According to the calculations in \cite{O} and \cite{A}  we have
isomorphism relations as  summarized in the following table:
\begin{center}
\small{
\begin{tabular}{|c|c|c|c|c|c|c| c|c|c|c|c|c|}\hline
$\begin{array}{c} \vspace{-.3cm} \\ 4A_{1} \\ \vspace{-.3cm}
\end{array}$ &$A_{4,1}$& ${A^a_{4,2}}$ & $A_{4,3}$ & $A _{4,4}$ &
${A^{a, b}_{4,5}}$ & ${A^{a, b}_{4,6}}$ & $A _{4,7}$ & $A _{4,8}$  &${A^b_{4,9}}$ & $A_{4,10}$ & ${A^a_{4,11}}$ & $A_{4,12}$ \\
\hline $\begin{array}{c} \vspace{-.3cm} \\ \mathfrak a_{4} \\
\vspace{-.3cm} \end{array}$ &$\mathfrak n_4$&$\mathfrak r_{4,a}$
& $\mathfrak r_{4,0}$ & $\mathfrak r_{4}$ & $\mathfrak r_{4,a,b}$
& $\mathfrak r_{4,a,b}'$ &
$\mathfrak h_{4}$ & $\mathfrak d _4$& $\mathfrak d_{4,1/1+b}$ & $\mathfrak d_{4,0}'$ &  $\mathfrak d_{4,a}'$ & $\mathfrak a\mathfrak f\mathfrak f(\mathbb{C})$\\
\hline
\end{tabular} }
\end{center}
\begin{center}
\small{
\begin{tabular}{|c|c|c|c|c|c|c| c|c|c|c|}\hline
$\begin{array}{c} \vspace{-.3cm} \\A_{2} \oplus A_{2} \\
\vspace{-.3cm}
\end{array}$ & $II\oplus R $ & $III\oplus R$ & $IV\oplus R$ &
$V\oplus R$ & $VI_{0}\oplus R$ & $VI_{a}\oplus R$ & $VII_{0}\oplus R$  &$VII_{a}\oplus R$  \\
&&&&&&$(a\neq 0,1)$&&$(a\neq 0)$ \\
 \hline
$\begin{array}{c} \vspace{-.3cm} \\ \mathfrak r_2\mathfrak r_2 \\
\vspace{-.3cm} \end{array}$ & $\mathfrak r\mathfrak h_{3}$ &
$\mathfrak r\mathfrak r_{3,0}$ & $\mathfrak r\mathfrak r_{3}$ &
$\mathfrak r\mathfrak r_{3,1}$ & $\mathfrak r\mathfrak r_{3,-1}$ &
$\mathfrak r\mathfrak r_{3,a}$ & $\mathfrak r\mathfrak r'_{3,0}$ & $\mathfrak r\mathfrak r'_{3,a}$ \\
\hline
\end{tabular} }
\end{center}
In this respect, one can see that in \cite{O} one complex
structure is obtained for Lie algebra $VII_{0}+R$ but according
to our calculation this Lie algebra has two non-equivalent
complex structures. For non-solvable Lie algebras $VIII+R$ and
$IX+R$ we obtain complex structures. In \cite{O}, two complex
structures are obtained for Lie algebra $A_{4,8}$ but here we
obtain four complex structures to this Lie algebra. Furthermore,
two complex structures are obtained for Lie algebra $A_{4,10}$ in
\cite{O}, but according to table 1 we obtain four complex
structures for this Lie algebra. Meanwhile, in \cite{O}, two
complex structures are obtained for Lie algebra $A_{4,12}$ but
here we obtain three complex structures for this Lie algebra. The
results for other Lie algebras are the same as \cite{O}.
\newpage
\vspace{10mm}
\begin{center}
 \begin{tabular}{|c|c|}
\multicolumn{2}{c}{\small{TABLE 1 : Complex structures on four dimensional real Lie algebras }}\\
\hline
\hline
Lie Algebra &  Complex Structures \\
\hline \hline
$III\oplus R $ & $Je_{1}=-e_{2} \hspace{2mm},\hspace{2mm} Je_{3}=e_{1}-e_{4} $  \\
\hline
$A_{2}\oplus A_{2}$ & $Je_{1}=-e_{2} \hspace{2mm},\hspace{2mm} Je_{3}=-e_{4} $ \\
\hline
$II \oplus R$ & $Je_{1}=-e_{4} \hspace{2mm},\hspace{2mm} Je_{2}=e_{3} $   \\
\hline
$ V \oplus R $ & $Je_{1}=-e_{4} \hspace{2mm},\hspace{2mm} Je_{2}=e_{3} $   \\
\hline
$VII_{0}\oplus R $ & $J_{1}e_{1}=-e_{2} \hspace{2mm},\hspace{2mm} J_{1}e_{3}=-e_{4} $\\
     & $J_{2}e_{1}=e_{2} \hspace{2mm},\hspace{2mm} J_{2}e_{3}=-e_{4} $  \\
\hline
$VII_{a}\oplus R$ & $J_{1}e_{1}=-e_{4} \hspace{2mm},\hspace{2mm} J_{1}e_{2}=e_{3} $\\
    \small$(a\neq0)$  & $J_{2}e_{1}=-e_{4} \hspace{2mm},\hspace{2mm} J_{2}e_{2}=-e_{3} $  \\
\hline
$VIII\oplus R $ & $Je_{1}=-e_{2} \hspace{2mm},\hspace{2mm} Je_{3}=pe_{3}-{(1+p^2)}e_{4}  \hspace{10mm} (p\in\mathbb{R}) $\\
\hline
$IX \oplus R $ & $Je_{1}=-e_{2} \hspace{2mm},\hspace{2mm} Je_{3}=pe_{3}-{(1+p^2)}e_{4}   \hspace{10mm} (p\in\mathbb{R}) $\\
\hline
 $A^1_{4,2}$ & $Je_{1}=e_{2} \hspace{2mm},\hspace{2mm} Je_{3}=e_{4} $  \\
\hline
 $A^{a,a}_{4,5}$ \small($-1\leq a < 1, a\neq 0)$ &$Je_{1}=-e_{4} \hspace{2mm},\hspace{2mm} Je_{2}=e_{3} $  \\
\hline
$A^{a,1}_{4,5}$ \small$(-1\leq a < 1, a\neq 0)$ & $Je_{1}=e_{3} \hspace{2mm},\hspace{2mm} Je_{2}=e_{4} $  \\
\hline
 $A^{1,1}_{4,5}$ & $Je_{1}=-e_{3} \hspace{2mm},\hspace{2mm} Je_{2}=-e_{4} $  \\
\hline
$A^{a,b}_{4,6}$ &  $J_{1}e_{1}=e_{4} \hspace{2mm},\hspace{2mm} J_{1}e_{2}=e_{3} $\\
   \small$(a\neq0, b\geq0)$   & $J_{2}e_{1}=e_{4} \hspace{2mm},\hspace{2mm} J_{2}e_{2}=-e_{3} $  \\
\hline
$A_{4,7}$ & $Je_{1}=e_{2} \hspace{2mm},\hspace{2mm} Je_{3}=-e_{4} $ \\
\hline
$A_{4,8}$ &  $J_{1}e_{1}=e_{2} \hspace{2mm},\hspace{2mm} J_{1}e_{3}=e_{4} $\\
    & $J_{2}e_{1}=-e_{3} \hspace{2mm},\hspace{2mm} J_{2}e_{2}=-e_{4} $  \\
     & $J_{3}e_{1}=e_{2} \hspace{2mm},\hspace{2mm} J_{3}e_{3}=e_{2}+e_{4} $  \\
      & $J_{4}e_{1}=-(e_{1}+2e_{3}) \hspace{2mm},\hspace{2mm} J_{4}e_{2}=-(e_{1}+e_{2})-2(e_{3}+e_{4}) $  \\
\hline $A^{b}_{4,9}$ &  $J_{1}e_{1}=-be_{2} \hspace{2mm},\hspace{2mm} J_{1}e_{3}=e_{4} $\\
\small$(0<\mid b \mid<1)$  & $J_{2}e_{1}=-e_{3} \hspace{2mm},\hspace{2mm} J_{2}e_{2}=-e_{4} $  \\
\hline
$A^{1}_{4,9}$ & $J_{1}e_{1}=-e_{3} \hspace{2mm},\hspace{2mm} J_{1}e_{2}=-e_{4} $\\
      & $J_{2}e_{1}=e_{4} \hspace{2mm},\hspace{2mm} J_{2}e_{2}=-e_{3}$  \\
      & $J_{3}e_{1}=-e_{4} \hspace{2mm},\hspace{2mm} J_{3}e_{2}=-e_{3} $  \\
\hline
$A^{0}_{4,9}$ & $Je_{1}=-e_{3} \hspace{2mm},\hspace{2mm} Je_{2}=-e_{4} $ \\
\hline
$A_{4,10}$ & $J_{1}e_{2}=-e_{1}+e_{3}\hspace{2mm},\hspace{2mm} J_{1}e_{3}=-e_{1}-e_{2}+e_{3}+e_{4}$\\
      & $ J_{2}e_{2}=e_{1}+e_{3}\hspace{2mm},\hspace{2mm} J_{2}e_{3}=-e_{1}-e_{2}-e_{3}+e_{4}$\\
      & $J_{3}e_{2}=-e_{1}-e_{3}\hspace{2mm},\hspace{2mm} J_{3}e_{3}=e_{1}+e_{2}+e_{3}-e_{4}$\\
     & $J_{4}e_{2}=-e_{1}-e_{3}\hspace{2mm},\hspace{2mm} J_{4}e_{3}=-e_{1}+e_{2}-e_{3}+e_{4}$\\
\hline
$A^{a}_{4,11}$ &  $J_{1}e_{1}=e_{4} \hspace{2mm},\hspace{2mm} J_{1}e_{2}=-e_{3} $\\
      \small$(a>0)$ & $J_{2}e_{1}=e_{4} \hspace{2mm},\hspace{2mm} J_{2}e_{2}=e_{3}$  \\
     & $J_{3}e_{1}=-e_{4} \hspace{2mm},\hspace{2mm} J_{3}e_{2}=-e_{3} $  \\
      & $J_{4}e_{1}=-e_{4} \hspace{2mm},\hspace{2mm} J_{4}e_{2}=e_{3} $  \\
\hline
$A_{4,12}$ & $J_{1}e_{1}=e_{2} \hspace{2mm},\hspace{2mm} J_{1}e_{3}=e_{4} $\\
      & $J_{2}e_{1}=-e_{2} \hspace{2mm},\hspace{2mm} J_{2}e_{3}=e_{4}$  \\
     & $J_{3}e_{1}=e_{4} \hspace{2mm},\hspace{2mm} J_{3}e_{2}=e_{3} $ \\
\hline
\end{tabular}
\end{center}
\newpage
\section{\bf BiHermitian  structures on four dimensional real Lie algebras}
{\bf Definition 4 \cite{G} \cite{SL} :}{\it If the complex
manifold M has two complex structures $J_{\pm}$ such that it is
Hermitian with respect to both complex structures. i.e.
\begin{equation}
J_{\pm}^{2}=-1\hspace{1mm},
\end{equation}
\begin{equation}
{N_{\mu\nu}}^{\kappa}(J_{\pm})=0\hspace{1mm},
\end{equation}
\begin{equation}
J_{\pm\mu}\hspace{0cm}^{\lambda}\hspace{1mm}
g_{\lambda\eta}\hspace{1mm}J_{\pm\nu}\hspace{0cm}^{\eta}=g_{\mu\nu},
\end{equation}
and furthermore if these complex structures be covariantly
constant with respect to certain connections $\Gamma^{\pm}$
\begin{equation}
\bigtriangledown^{\pm}_{\mu}\hspace{2mm}J_{\pm\nu}\hspace{0cm}^{\lambda}\equiv{{J_{\pm\nu}\hspace{0cm}^{\lambda}}_{,\mu}}
+{\Gamma^{\pm}_{\mu\rho}}^{\lambda}J_{\pm\nu}\hspace{0cm}^{\rho}-{\Gamma^{\pm}_{\mu\nu}}^{\rho}J_{\pm\rho}\hspace{0cm}^{\lambda}=0,
\end{equation}
with
\begin{equation}
{\Gamma^{\pm}_{\mu\nu}}^{\lambda}={\Gamma_{\mu\nu}}^{\lambda} \pm
{T_{\mu\nu}}^{\lambda}\hspace{2mm},\hspace{2mm}
{T_{\mu\nu}}^{\lambda}=H_{\mu\nu\eta}g^{\eta\lambda},
\end{equation}
then it is said that $M$ has biHermitian structure, shown by
$(M,g,J_{\pm})$.}
\vspace{.5cm}\\
In the above definition $g_{\mu\nu},{\Gamma_{\mu\nu}}^{\lambda}$
and $H_{\mu\nu\eta}$ are metric , Christoffel connection and
antisymmetric tensors on M respectively. Using $(22)$, the
integrability condition $(20)$ may be rewritten in the following
form \cite{SL}:
\begin{equation}
H_{\delta\nu\lambda}= J_{\pm\delta}\hspace{0cm}^{\sigma}
J_{\pm\nu}\hspace{0cm}^{\rho}H_{\sigma\rho\lambda}+J_{\pm\delta}\hspace{0cm}^{\rho}
J_{\pm\lambda}\hspace{0cm}^{\sigma}H_{\sigma\rho\nu}+J_{\pm\upsilon}\hspace{0cm}^{\sigma}
J_{\pm\lambda}\hspace{0cm}^{\rho}H_{\sigma\rho\delta}\hspace{1mm}.
\end{equation}
Furthermore, by introducing the K\"{a}hler forms
\begin{equation}
\omega_{\pm\mu\nu}\equiv\ g_{\mu\nu}
J_{\pm\nu}\hspace{0cm}^{\lambda},
\end{equation}
and by use of $(22)$ one can find
\begin{equation}
(d\omega_{\pm})_{\rho\mu\nu}=\pm(H_{\sigma\rho\mu}J^{\sigma}_{\pm\nu}+H_{\sigma\mu\nu}J^{\sigma}_{\pm\rho}+H_{\sigma\nu\rho}J^{\sigma}_{\pm\mu}),
\end{equation}
where
\begin{equation}
(d\omega_{\pm})_{\lambda\sigma\gamma}=\frac{1}{2}(\partial_{\lambda}\omega_{\pm\sigma\gamma}+\partial_{\sigma}\omega_{\pm\gamma\lambda}+\partial_{\gamma}\omega_{\pm\lambda\sigma}).
\end{equation}
Finally, using $(25)$ and $(26)$ one can find
\begin{equation}
H_{\mu\nu\rho}=-J^{\lambda}_{+\mu}J^{\sigma}_{+\nu}J^{\gamma}_{+\rho}(d\omega_{+})_{\lambda\sigma\gamma}=-J^{\lambda}_{-\mu}J^{\sigma}_{-\nu}J^{\gamma}_{-\rho}(d\omega_{-})_{\lambda\sigma\gamma}.
\end{equation}
In this respect, the target manifold $(M,g,J_{\pm})$ is said to
have biHermitian structure if two Hermitian complex structures
$J_{\pm}$ satisfy the relation $(28)$ (i.e. relation between
$(J_{+},\omega_{+})$ and $(J_{-},\omega_{-})$)which defines the
torsion H. Now, for the case where M is a Lie group G, similar to
the process presented in section 2 one can transform relations
$(19)-(22)$ and $(24)$ to the algebraic relation using the
relations $(6),(7)$ and the following relations:
\begin{equation}
g_{\alpha\beta}=L_{\alpha}\hspace{0cm}^{\mu}L_{\beta}\hspace{0cm}^{\nu}g_{\mu\nu}=R_{\alpha}\hspace{0cm}^{\mu}R_{\beta}\hspace{0cm}^{\nu}g_{\mu\nu}\hspace{2mm},\hspace{2mm}
g_{\mu\nu}=L^{\alpha}\hspace{0cm}_{\mu}L^{\beta}\hspace{0cm}_{\nu}g_{\alpha\beta}=R^{\alpha}\hspace{0cm}_{\mu}R^{\beta}\hspace{0cm}_{\nu}g_{\alpha\beta},
\end{equation}

\begin{equation}
H_{\mu\nu\rho}=\frac{1}{2}
L^{\alpha}\hspace{0cm}_{\mu}L^{\beta}\hspace{0cm}_{\nu}L^{\gamma}\hspace{0cm}_{\rho}H_{\alpha\beta\gamma}=\frac{1}{2}
R^{\alpha}\hspace{0cm}_{\mu}R^{\beta}\hspace{0cm}_{\nu}R^{\gamma}\hspace{0cm}_{\rho}H_{\alpha\beta\gamma},
\end{equation}

\begin{equation}
J_{+\mu}\hspace{0cm}^{\nu}=R^{\alpha}\hspace{0cm}_{\mu}{J_{\alpha}}\hspace{0cm}^{\beta}R_{\beta}\hspace{0cm}^{\nu}\hspace{1mm},\hspace{1mm}J_{-\mu}\hspace{0cm}^{\nu}=L^{\alpha}\hspace{0cm}_{\mu}{J_{\alpha}}\hspace{0cm}^{\beta}L_{\beta}\hspace{0cm}^{\nu},
\end{equation}
where $L^{\alpha}\hspace{0cm}_{\mu}(
R^{\alpha}\hspace{0cm}_{\mu})$ and
$L_{\beta}\hspace{0cm}^{\nu}(R_{\beta}\hspace{0cm}^{\nu})$ are
left(right) invariant vierbeins and their inverses respectively.
Now using these relations , $(21)$ and $(24)$ transform to the
following matrix relations:
\begin{equation}
J\hspace{1mm}g\hspace{1mm}J^{t}=g,
\end{equation}
\begin{equation}
H_{\alpha}= J (H_{\beta} {J_{\alpha}}\hspace{0cm}^{\beta}) +
JH_{\alpha}J^{t}+(H_{\beta} {J_{\alpha}}\hspace{0cm}^{\beta})
J^{t},
\end{equation}
\vspace{1cm}\\

where
$(H_{\alpha})_{\beta\gamma}={H_{\alpha}}\hspace{0cm}_{\beta\gamma}$.
Furthermore, using the following relations \cite{LI}
{\footnote{Note that in these relations all algebraic (target)
indices are lowered and raised by $g_{\alpha\beta}(g_{\mu\nu})$;
furthermore, these indices transform into each other by
$L_{\alpha}\hspace{0cm}^{\mu}(R_{\alpha}\hspace{0cm}^{\mu})$ or
$L^{\alpha}\hspace{0cm}_{\mu}(R^{\alpha}\hspace{0cm}_{\mu})$. The
symmetrization notation have the following form:
$f_{\alpha}\hspace{0cm}^{(\rho\mu)}=f_{\alpha}\hspace{0cm}^{\rho\mu}+f_{\alpha}\hspace{0cm}^{\mu\rho}$.}
\begin{equation}
\bigtriangledown^{\rho}{L_{\alpha}}\hspace{0cm}^{\mu}=-\frac{1}{2}(f_{\alpha}\hspace{0cm}^{(\rho\mu)}+{f^{\rho\mu}}\hspace{0cm}_{\alpha}+T_{\alpha}\hspace{0cm}^{(\rho\mu)}+{T^{\rho\mu}}\hspace{0cm}_{\alpha}
+{L_{\beta}}\hspace{0cm}^{\rho}{L_{\gamma}}\hspace{0cm}^{\mu}{\bigtriangledown}_{\alpha}g^{\beta\gamma}+L^{\beta\rho}\bigtriangledown^{\mu}g_{\alpha\beta}-L^{\beta\mu}\bigtriangledown^{\rho}g_{\alpha\beta}),
\end{equation}
\begin{equation}
\bigtriangledown^{\rho}{R_{\alpha}}\hspace{0cm}^{\mu}=-\frac{1}{2}(-f_{\alpha}\hspace{0cm}^{(\rho\mu)}-{f^{\rho\mu}}\hspace{0cm}_{\alpha}+T_{\alpha}\hspace{0cm}^{(\rho\mu)}+{T^{\rho\mu}}\hspace{0cm}_{\alpha}
+{R_{\beta}}\hspace{0cm}^{\rho}{R_{\gamma}}\hspace{0cm}^{\mu}{\bigtriangledown}_{\alpha}g^{\beta\gamma}+R^{\beta\rho}\bigtriangledown^{\mu}g_{\alpha\beta}-R^{\beta\mu}\bigtriangledown^{\rho}g_{\alpha\beta}),
\end{equation}
and assuming that $g_{\alpha\beta}$ are coordinate independent;
the relation $(22)$ transforms to the following algebraic
relation{\footnote{This relation can also be obtained from
algebraic form of $(28)$.}}:
\begin{equation}
J(H_{\alpha}-\chi_{\alpha}g) =(J(H_{\alpha}-\chi_{\alpha}g))^{t}.
\end{equation}
Note that the metric $g_{\alpha\beta}$ is the ad invariant metric
on Lie algebras ${\bf g}$ .i.e.\hspace{.5mm}we have
\begin{equation}
\langle{X_{\alpha},X_{\beta}}\rangle=g_{\alpha\beta},
\end{equation}
\begin{equation}
\langle{X_{\alpha},[X_{\beta},X_{\gamma}]}\rangle=\langle{[X_{\alpha},X_{\beta}],X_{\gamma}}\rangle,
\end{equation}
or in matrix notation we have {\footnote{For semisimple Lie
algebras, the Killing form is one of nondegenerate solutions for
this equation and other nondegenerate solution may exist. For
nonsemisimple Lie algebras the Killing tensor degenerates and
there may exist nondegenerate solution for $(39)$.} }
\begin{equation}
\chi_{\alpha}g=-(\chi_{\alpha}g)^{t}.
\end{equation}
 Now, one can obtain biHermitian structures on Lie
algebras by solving relations $(9),(12),(32),(33),(36)$ and $(39)$
simultaneously. These relations can be applied on the Lie algebra
as a definition of algebraic biHermitian structure on
{\bf g}\hspace{1mm};\\

{\bf Definition 5:} {\it If there exist endomorphism $J:\bf g
\rightarrow \bf g$ of Lie algebra with ad invariant metric g and
antisymmetric bilinear map $H:\bf g\otimes \bf g \rightarrow \bf
g$ such that the relations
$(9),(12),(32),(33),(36)$ and $(39)$ are satisfied, then we have biHermitian structure $(J,g,H)$ on {\bf g}.}\\

Note that relation $(33)$ is equivalent to matrix relation of
integrability condition i.e. relation $(12)$ . For this reason,
first it is better to obtain algebraic complex structures J, then
solve relations $(32),(36)$ and $(39)$ and finally check them in
$(33)$. We do this for real four dimensional Lie algebras using
Maple. Note that similar to complex structures, in order to obtain
non-equivalent biHermitian
structures we suggest the following equivalent relations. \\

{\it{\bf Definition 6} : Two biHermitian structures $(J,g,H)$ and
$(J^{'},g^{'},H^{'})$ of Lie algebras {\bf g} are equivalent if
there exists an element A of automorphism group of Lie algebra
{\bf g} (Auto {\bf g}) such that:
\begin{equation}
J^{'}=A J A^{-1},
\end{equation}
\begin{equation}
g^{'}=A g A^{t},
\end{equation}
\begin{equation}
H^{'}_{\alpha}=A(H_{\beta} A_{\alpha}\hspace{0cm}^{\beta})A^{t}.
\end{equation}}

These relations are equivalent relations and are satisfied in the
equivalent conditions. Note that if
$f_{\beta\gamma}\hspace{0cm}^{\alpha}=H_{\delta\beta\gamma}g^{\delta\alpha}$
or $H$ is isomorphic with $f$, i.e. if there exists isomorphism
matrix $C$ such that
\begin{equation}
C Y^{\alpha} C^{t}= \tilde{Y}^{\beta }
C_{\beta}\hspace{0cm}^{\alpha},
\end{equation}
where
$(Y^{\alpha})_{\beta\gamma}=-f_{\beta\gamma}\hspace{0cm}^{\alpha}$
and
$({\tilde{Y}}^{\alpha})_{\beta\gamma}=-H_{\delta\beta\gamma}g^{\delta\alpha}$;
then $(J,g,H)$ shows the Manin triple structure on ${\bf
g}$\cite{L}.
 In this way biHermitian structures on real four dimensional Lie
algebras can be classified as table 2 . Note that according to the
table for Lie algebra $A_{4,8}$, we have two non-equivalent
biHermitian structure $(J,g,H)$ where the second biHermitian
structure  shows the Manin triple structure of $A_{4,8}$ \cite{L}
(i.e. $A_{4,8}$ is a Manin triple of two dimensional Lie
bialgebras (type B and semiabelian)\cite{Sn}) for the following
values of parameters:
$$
c_1=c_2=c_3=c_4=c_5=c_6=c_7=c_9=c_{10}=c_{11}=c_{13}=c_{14}=0\hspace{1mm},\hspace{1mm}
c_{12}=c_{15}=-1.
$$
 For Lie algebras $VIII \oplus R$, there is one biHermitian
structure where this structure, for the values $$
d_1=d_2=d_4=d_5=d_7=d_8=d_{10}=d_{11}=d_{12}=d_{14}=d_{15}=d_{16}=0\hspace{1mm},\hspace{1mm}
d_9\neq 0\hspace{1mm},\hspace{2mm}d_3=-d_6=\alpha,
$$
is isomorphic with two dimensional Lie bialgebra type A
\cite{Sn}. There exists one biHermitian structure for Lie algebra
$IX \oplus R$. The results are given in table 2 {\footnote{Note
that results of table 1 are solutions of the equations $(9)$ and
$(12)$. But the results of table 2 are solutions of $(9)$,$(12)$
and $(32)$,$(33)$,$(36)$,$(39)$ so the results of table 2 can
also be solutions of these equations which must be consistent
with H and  $g$ .}}. Note that the isomorphism relation
$(43)$(i.e. the biHermitian structures which show Manin triple)
are independent of the choice of special biHermitian structures
from equivalent class of biHermitian structures. In this way if
relation $(43)$ holds ; then by
$\tilde{Y}^{'\alpha}=-H^{'}_{\delta} g^{'\delta\alpha}$ and using
relations $(42)$ and $(43)$ one can show that
\begin{equation}
(AC) Y^{\gamma} (AC)^{t}=\tilde{Y}^{'\alpha}
(AC)_{\alpha}\hspace{0cm}^{\gamma}.
\end{equation}
\begin{center}
\begin{tabular}{|c|c|c|c|}
\multicolumn{4}{c}{TABLE 2 : biHermitian structures on four
dimensional real Lie algebras  }\\
\hline \hline Lie Algebra & Complex Structures & $g$ &
antisymmetric tensor \\
\hline
\hline
 $A_{4,8} $&&&
 $H_{1}=\left(
\begin{array}{cccc}
  0 & b_{1} & 1 & -b_{2} \\
  -b_{1} & 0 & b_{2} & 1 \\
  -1 & -b_{2} & 0 & b_{3} \\
  b_{2} & -1 & -b_{3} & 0 \\
\end{array}
\right)$\\&$J=\left(
\begin{array}{cccc}
  0 & 1 & 0 & 0 \\
  -1 & 0 & 0 & 0 \\
  0 & 0 & 0 & 1 \\
 0 & 0 & -1 & 0 \\
\end{array}
\right)$ &$g=\left(
\begin{array}{cccc}
  0 & 0 & 0 & 1 \\
  0 & 0 & -1 & 0 \\
  0 & -1 & 0 & 0 \\
 1 & 0 & 0 & 0 \\
\end{array}
\right)$&
$H_{2}=\left(
\begin{array}{cccc}
  0 & b_{4} & b_{5} & -b_{6} \\
  -b_{4} & 0 & b_{6} & b_{5} \\
  -b_{5} & -b_{6} & 0 & b_{7} \\
  b_{6} & -b_{5} & -b_{7} & 0 \\
\end{array}
\right)$\\
&&&$H_{3}=\left(
\begin{array}{cccc}
  0 & b_{8} & b_{9} & -b_{10} \\
  -b_{8} & 0 & b_{10} & 1+b_{9} \\
  -b_{9} & -b_{10} & 0 & b_{11} \\
  b_{10} & -1-b_{9} & -b_{11} & 0 \\
\end{array}
\right)$\\&&&$H_{4}=\left(
\begin{array}{cccc}
  0 & b_{12} & b_{13} & -1-b_{14} \\
  -b_{12} & 0 & b_{14} & b_{13} \\
  -b_{13} & -b_{14} & 0 & b_{15} \\
  1+b_{14} & -b_{13} & -b_{15} & 0 \\
\end{array}
\right)$\\
 \cline {2-4}
 &&& $H_{1}=\left(
\begin{array}{cccc}
  0 & c_{1} & c_{2} & c_{3} \\
  -c_{1} & 0 & c_{3} & c_{4} \\
  -c_{2} & -c_{3} & 0 & c_{1} \\
  -c_{3} & -c_{4} & -c_{1} & 0 \\
\end{array}
\right)$\\&$J=\left(
\begin{array}{cccc}
  0 & 0 & -1 & 0 \\
  0 & 0 & 0 & -1 \\
  1 & 0 & 0 & 0 \\
 0 & 1 & 0 & 0 \\
\end{array}
\right)$&$g=\left(
\begin{array}{cccc}
  0 & 0 & 0 & -1 \\
  0 & 0 & 1 & 0 \\
  0 & 1 & 0 & 0 \\
 -1 & 0 & 0 & 0 \\
\end{array}
\right)$&$H_{2}=\left(
\begin{array}{cccc}
  0 & c_{5}-1 & c_{6} & c_{7} \\
  -c_{5}+1 & 0 & c_{7} & c_{8} \\
  -c_{6} & -c_{7} & 0 & c_{5} \\
  -c_{7} & -c_{8} & -c_{5} & 0 \\
\end{array}
\right)$\\
&&&$H_{3}=\left(
\begin{array}{cccc}
  0 & c_{9} & c_{10} & c_{11} \\
  -c_{9} & 0 & c_{11} & c_{12} \\
  -c_{10} & -c_{11} & 0 & c_{9} \\
  -c_{11} & -c_{12} & -c_{9} & 0 \\
\end{array}
\right)$\\&&&$H_{4}=\left(
\begin{array}{cccc}
  0 & c_{13} & c_{14} & c_{15} \\
  -c_{13} & 0 & 1+c_{15} & c_{16} \\
  -c_{14} & -1-c_{15} & 0 & c_{13} \\
  -c_{15} & -c_{16} & -c_{13} & 0 \\
\end{array}
\right)$\\
\hline
\end{tabular}
\end{center}
\begin{center}
\begin{tabular}{|c|c|c|c|}
\multicolumn{4}{c}{TABLE 2 : biHermitian structures on four
dimensional real Lie algebras }\\
\hline
 \hline
Lie Algebra & Complex Structures & $g$ & antisymmetric tensor \\
\hline
\hline
 $VIII \oplus R$&&& $H_{1}=\left(
\begin{array}{cccc}
  0 & d_{1} & d_{2} & -d_{3}+\alpha \\
  -d_{1} & 0 & d_{3} & d_{2} \\
  -d_{2} & -d_{3} & 0 & d_{4} \\
  d_{3}-\alpha & -d_{2} & -d_{4} & 0 \\
\end{array}
\right)$\\&$J=\left(
\begin{array}{cccc}
  0 & -1 & 0 & 0 \\
  1 & 0 & 0 & 0 \\
  0 & 0 & 0 & -1 \\
 0 & 0 & 1 & 0 \\
\end{array}
\right)$ &$g=\left(
\begin{array}{cccc}
  -\alpha & 0 & 0 & 0 \\
  0 & -\alpha & 0 & 0 \\
  0 & 0 & \alpha & 0 \\
 0 & 0 & 0 & \alpha \\
\end{array}
\right)$&$H_{2}=\left(
\begin{array}{cccc}
  0 & d_{5} & d_{6} & -d_{7} \\
  -d_{5} & 0 & d_{7} & d_{6}+\alpha \\
  -d_{6} & -d_{7} & 0 & d_{8} \\
  d_{7} & -d_{6}-\alpha & -d_{8} & 0 \\
\end{array}
\right)$\\
&&$\alpha \in R-\{0\}$&$H_{3}=\left(
\begin{array}{cccc}
  0 & d_{9} & d_{10} & -d_{11} \\
  -d_{9} & 0 & d_{11} & d_{10} \\
  -d_{10} & -d_{11} & 0 & d_{12} \\
  d_{11} & -d_{10} & -d_{12} & 0 \\
\end{array}
\right)$\\&&&$H_{4}=\left(
\begin{array}{cccc}
  0 & d_{13} & d_{14} & -d_{15} \\
  -d_{13} & 0 & d_{15} & d_{14} \\
  -d_{14} & -d_{15} & 0 & d_{16} \\
 d_{15} & -d_{14} & -d_{16} & 0 \\
\end{array}
\right)$\\
\hline $IX \oplus R$&&&
$H_{1}=\left(
\begin{array}{cccc}
  0 & 0 & f_{1} & -f_{2}-\beta \\
  0 & 0 & f_{2} & f_{1} \\
  -f_{1} & -f_{2} & 0 & f_{3} \\
  f_{2}+\beta & -f_{1} & -f_{3} & 0 \\
\end{array}
\right)$\\&$J=\left(
\begin{array}{cccc}
  0 & -1 & 0 & 0 \\
  1 & 0 & 0 & 0 \\
  0 & 0 & 0 & -1 \\
 0 & 0 & 1 & 0 \\
\end{array}
\right)$ &$g=\left(
\begin{array}{cccc}
  \beta & 0 & 0 & 0 \\
  0 & \beta & 0 & 0 \\
  0 & 0 & \beta & 0 \\
 0 & 0 & 0 & \beta \\
\end{array}
\right)$&$H_{2}=\left(
\begin{array}{cccc}
  0 & f_{4} & f_{5} & -f_{6} \\
  -f_{4} & 0 & f_{6} & f_{5}-\beta \\
  -f_{5} & -f_{6} & 0 & f_{7} \\
  f_{6} & -f_{5}+\beta & -f_{7} & 0 \\
\end{array}
\right)$\\
&&$\beta \in R-\{0\}$&$H_{3}=\left(
\begin{array}{cccc}
  0 & f_{8} & f_{9} & -f_{10} \\
  -f_{8} & 0 & f_{10} & f_{9} \\
  -f_{9} & -f_{10} & 0 & f_{11} \\
  f_{10} & -f_{9} & -f_{11} & 0 \\
\end{array}
\right)$\\&&&$H_{4}=\left(
\begin{array}{cccc}
  0 & f_{12} & f_{13} & -f_{14} \\
  -f_{12} & 0 & f_{14} & f_{13} \\
  -f_{13} & -f_{14} & 0 & f_{15} \\
  f_{14} & -f_{13} & -f_{15} & 0 \\
\end{array}
\right)$\\
\hline
\end{tabular}
\end{center}
Note that $b_{i}, c_{i}, d_{i}, f_{i}$ are all real parameters.
\vspace{1cm}
\section{\bf Conclusion}
We offered a new method for calculation of complex and biHermitian
structures on low dimensional Lie algebras. By this method, we
obtain complex and biHermitian structures on real four
dimensional Lie algebras. In this manner, one can obtain these
structures on Lie groups using vierbeins . Some biHermitian
structures on real four dimensional Lie algebras are equivalent to
Manin triple structure obtained in \cite{Sn}. One can use these
methods for obtaining complex and biHermitian structures on real
six dimensional Lie algebras \cite{RS}. We also apply this method
for calculation of generalized complex structures on four
dimensional real Lie algebras \cite{SRS}.
\vspace{3mm}\\

 {\bf Acknowledgments}

\vspace{3mm} We would like to thank F. Darabi, Sh. Mogadassi and
Z. Haghpanah for carefully reading the manuscript and useful
comments.

\vspace{5mm}
\newpage
\hspace{-6.5mm}{\bf{{Appendix\hspace{2mm}}}} \vspace{3mm}\\ {\bf
Real four dimensional Lie algebras and their automorphisms groups
\cite{P},\cite{PaP} } \vspace{2mm}
\begin{center}
\begin{tabular}{|c|c|c|}
\multicolumn{3}{c}{TABLE A: Classifications of four dimensional real Lie algebras  }\\
  \hline
  \hline
Lie Algebra & Non Vanishing & Automorphisms group \\&Structure Constants&\\
\hline
\hline
$4A_{1}$ &  & $$\\
\hline
\vspace{-4mm}
$III\oplus R\cong\left(A_{2}\oplus
2A_{1}\right)$ & $f^{2}_{12}=-1,f^{3}_{12}=-1$ & $\left(
\begin{array}{cccc}
  1 & a_{1} & a_{2} & a_{3} \\
 0 & a_{5} & a_{4} & -a_{6} \\
  0 & a_{4} & a_{5} & a_{6} \\
   0 & a_{7} & -a_{7} & a_{8} \\
\end{array}
\right)$  \\
 & $,f^{2}_{31}=1,f^{3}_{31}=1$\\
 \hline
$2A_{2}$ & $f^{2}_{12}=1,f^{4}_{34}=1$ & $\left(
\begin{array}{cccc}
  1 & a_{1} & 0 & 0 \\
  0 & a_{2} & 0 & 0 \\
  0 & 0 & 1 & a_{3} \\
  0 & 0 & 0 & a_{4} \\
\end{array}
\right)$\\
\hline
  $II\oplus R\cong\left(A_{3,1}\oplus A_{1}\right)$ & $f^{1}_{23}=1$ & $\left(
\begin{array}{cccc}
  a_{2}a_{7}-a_{3}a_{6} & 0 & 0 & 0 \\
  a_{1} & a_{2} & a_{3} & a_{4} \\
  a_{5} & a_{6} & a_{7} & a_{8} \\
  a_{9} & 0 & 0 & a_{10} \\
\end{array}
\right)$  \\
  \hline
$IV\oplus R\cong\left(A_{3,2}\oplus A_{1}\right)$ &
$f^{2}_{12}=-1,f^{3}_{12}=1,f^{3}_{13}=-1$ & $\left(
\begin{array}{cccc}
 1 & a_{1} & a_{2} & a_{3} \\
  0 & a_{4} & a_{5} & 0 \\
  0 & 0 & a_{4} & 0 \\
  0 & 0 & 0 & a_{6} \\
\end{array}
\right)$  \\
\hline $V\oplus R\cong\left(A_{3,3}\oplus A_{1}\right)$ &
$f^{2}_{12}=-1,f^{3}_{13}=-1$ & $\left(
\begin{array}{cccc}
 1 & a_{1} & a_{2} & a_{3} \\
  0 & a_{4} & a_{5} & 0 \\
  0 & a_{6} & a_{7} & 0 \\
  0 & 0 & 0 & a_{8} \\
\end{array}
\right)$  \\
\hline $VI_{0}\oplus R\cong\left(A_{3,4}\oplus A_{1}\right)$ &
$f^{2}_{13}=1,f^{1}_{23}=1$ & $\left(
\begin{array}{cccc}
  a_{2} & a_{1} & 0 & 0 \\
 a_{1} & a_{2} & 0 & 0 \\
  a_{3} & a_{4} & 1 & a_{5} \\
  0 & 0 & 0 & a_{6} \\
\end{array}
\right)$\\
\hline
 \vspace{-4mm}
$VI_{a}\oplus R\cong\left(A^{a}_{3,5}\oplus A_{1}\right)$ &
$f^{2}_{12}=-a,f^{3}_{12}=-1$ & $\left(
\begin{array}{cccc}
 1 & a_{1} & a_{2} & a_{3} \\
  0 & a_{5} & a_{4} & 0 \\
  0 & a_{4} & a_{5} & 0 \\
  0 & 0 & 0 & a_{6} \\
\end{array}
\right)$\\
&  $,f^{2}_{31}=1,f^{3}_{31}=a$\\
 \hline
 $VII_{0}\oplus R\cong\left(A_{3,6}\oplus A_{1}\right)$ & $f^{1}_{23}=1,f^{2}_{13}=-1$ & $\left(
\begin{array}{cccc}
  a_{2} & -a_{1} & 0 & 0 \\
  a_{1} & a_{2} & 0 & 0 \\
  a_{3} & a_{4} & 1 & a_{5} \\
 0 & 0 & 0 & a_{6} \\
\end{array}
\right)$  \\
\hline
 \vspace{-4mm}
 $VII_{a}\oplus R\cong\left(A^{a}_{3,7}\oplus A_{1}\right)$ & $f^{2}_{31}=1,f^{3}_{31}=a$ & $\left(
\begin{array}{cccc}
  1 & a_{1} & a_{2} & a_{3} \\
  0 & a_{5} & -a_{4} & 0 \\
  0 & a_{4} & a_{5} & 0 \\
 0 & 0 & 0 & a_{6} \\
\end{array}
\right)$  \\
&  $,f^{2}_{12}=-a,f^{3}_{12}=1$\\
\hline
\end{tabular}
\end{center}
\newpage
\vspace{10mm}
\begin{center}
\begin{tabular}{|c|c|c|}
\multicolumn{3}{c}{TABLE A: Classifications of four dimensional real Lie algebras  }\\
  \hline
  \hline
  Lie Algebra & Non Vanishing & Automorphisms group \\&Structure Constants&\\
  \hline
  \hline
  $VIII\oplus R\cong\left(A_{3,8}\oplus A_{1}\right)$ & $f^{2}_{31}=1,f^{3}_{12}=-1,f^{1}_{23}=1$ & $\Lambda_{1}$  \\
  \hline
  $IX\oplus R\cong\left(A_{3,9}\oplus A_{1}\right)$ & $f^{2}_{31}=1,f^{3}_{12}=1,f^{1}_{23}=1$ & $\Lambda_{2}$\\
 \hline
  $A_{4,1}$ & $f^{1}_{24}=1,f^{2}_{34}=1$ & $\left(
\begin{array}{cccc}
  a^{2}_{7}a_{3} & 0 & 0 & 0 \\
  a_{2}a_{7} & a_{3}a_{7} & 0 & 0 \\
  a_{1} & a_{2} & a_{3} & 0 \\
  a_{4} & a_{5} & a_{6} & a_{7} \\
\end{array}
\right)$  \\
  \hline
\vspace{-4mm}
 $A^{a}_{4,2}$ & $f^{1}_{14}=a,f^{2}_{24}=1$ &
$\left(
\begin{array}{cccc}
  a_{1} & 0 & 0 & 0 \\
  0 & a_{3} & 0 & 0 \\
  0 & a_{2} & a_{3} & 0 \\
  a_{4} & a_{5} & a_{6} & 1 \\
\end{array}
\right)$  \\
&  $,f^{2}_{34}=1,f^{3}_{34}=1$\\
 \hline
\vspace{-4mm}
 $A^{1}_{4,2}$&
$f^{1}_{14}=1,f^{2}_{24}=1$ & $\left(
\begin{array}{cccc}
  a_{1} & a_{2} & 0 & 0 \\
  0 & a_{5} & 0 & 0 \\
  a_{3} & a_{4} & a_{5} & 0 \\
  a_{6} & a_{7} & a_{8} & 1 \\
\end{array}
\right)$  \\
&  $,f^{2}_{34}=1,f^{3}_{34}=1$\\
 \hline
 $A_{4,3}$ & $f^{1}_{14}=1,f^{2}_{34}=1$ & $\left(
\begin{array}{cccc}
  a_{1} & 0 & 0 & 0 \\
 0 & a_{2} & 0 & 0 \\
  0 & a_{3} & a_{2} & 0 \\
  a_{4} & a_{5} & a_{6} & 1 \\
\end{array}
\right)$  \\
\hline
 \vspace{-4mm}
 $A_{4,4}$ &
$f^{1}_{14}=1,f^{1}_{24}=1,f^{2}_{24}=1$ & $\left(
\begin{array}{cccc}
  a_{3} & 0 & 0 & 0 \\
  a_{2} & a_{3} & 0 & 0 \\
  a_{1} & a_{2} & a_{3} & 0 \\
  a_{4} & a_{5} & a_{6} & 1 \\
\end{array}
\right)$\\
& $,f^{2}_{34}=1,f^{3}_{34}=1$\\
 \hline
\vspace{-4mm}
 $A^{a,b}_{4,5}$ &
$f^{1}_{14}=1,f^{2}_{24}=a$ & $\left(
\begin{array}{cccc}
  a_{1} & 0 & 0 & 0 \\
  0 & a_{2} & 0 & 0 \\
  0 & 0 & a_{3} & 0 \\
  a_{4} & a_{5} & a_{6} & 1 \\
\end{array}
\right)$\\
&  $,f^{3}_{34}=b$\\
 \hline
\vspace{-4mm}
 $A^{a,a}_{4,5}$ & $f^{1}_{14}=1,f^{2}_{24}=a$ & $\left(
\begin{array}{cccc}
  a_{1} & 0 & 0 & 0 \\
  0 & a_{2} & a_{3} & 0 \\
  0 & a_{4} & a_{5} & 0 \\
  a_{6} & a_{7} & a_{8} & 1 \\
\end{array}
\right)$  \\
&  $,f^{3}_{34}=a$\\
 \hline
\vspace{-4mm}
 $A^{a,1}_{4,5}$ &
$f^{1}_{14}=1,f^{2}_{24}=a$& $\left(
\begin{array}{cccc}
  a_{1} & 0 & a_{2} & 0 \\
  0 & a_{3} & 0 & 0 \\
  a_{4} & 0 & a_{5} & 0 \\
  a_{6} & a_{7} & a_{8} & 1 \\
\end{array}
\right)$  \\
& $,f^{3}_{34}=1$\\
 \hline
 \vspace{-4mm}
 $A^{1,1}_{4,5}$ & $f^{1}_{14}=1,f^{2}_{24}=1$ & $\left(
\begin{array}{cccc}
  a_{1} & a_{2} & a_{3} & 0 \\
  a_{4} & a_{5} & a_{6} & 0 \\
  a_{7} & a_{8} & a_{9} & 0 \\
  a_{10} & a_{11} & a_{12} & 1 \\
\end{array}
\right)$  \\
&  $,f^{3}_{34}=1$\\
 \hline
\end{tabular}
\end{center}
\newpage
\vspace{10mm}
\begin{center}
\begin{tabular}{|c|c|c|}
\multicolumn{3}{c}{TABLE A: Classifications of four dimensional real Lie algebras }\\
  \hline
  \hline
  Lie Algebra & Non Vanishing & Automorphisms group \\&Structure Constants&\\
  \hline
  \hline
\vspace{-4mm}
  $A^{a,b}_{4,6}$ & $f^{1}_{14}=a,f^{2}_{24}=b,f^{3}_{24}=-1$ & $\left(
\begin{array}{cccc}
  a_{1} & 0 & 0 & 0 \\
  0 & a_{3} & -a_{2} & 0 \\
 0 & a_{2} & a_{3} & 0 \\
  a_{4} & a_{5} & a_{6} & 1 \\
\end{array}
\right)$  \\
& $,f^{2}_{34}=1,f^{3}_{34}=b$\\
\hline
\vspace{-4mm}
  $A_{4,7}$ & $f^{1}_{14}=2,f^{2}_{24}=1,f^{2}_{34}=1$ & $\left(
\begin{array}{cccc}
  a^2_{2} & 0 & 0 & 0 \\
  -a_{2}a_{5} & a_{2} & 0 & 0 \\
  -a_{2}a_{5}+a_{2}a_{4}-a_{1}a_{5} & a_{1} & a_{2} & 0 \\
 a_{3} & a_{4} & a_{5} & 1 \\
\end{array}
\right)$\\
& $,f^{3}_{34}=1,f^{1}_{23}=1$\\
 \hline
\vspace{-5mm}
  $A_{4,8}$& $f^{2}_{24}=1,f^{3}_{34}=-1$ & $\left(
\begin{array}{cccc}
  a_{1}a_{2} & 0 & 0 & 0 \\
  a_{1}a_{5} & a_{1} & 0 & 0 \\
  a_{2}a_{4} & 0 & a_{2} & 0 \\
a_{3} & a_{4} & a_{5} & 1 \\
\end{array}
\right)$  \\
& $,f^{1}_{23}=1$\\
  \hline
\vspace{-5mm}
 $A^{b}_{4,9}$ &
$f^{1}_{14}={1+b},f^{2}_{24}=1$ & $\left(
\begin{array}{cccc}
  a_{1}a_{2} & 0 & 0 & 0 \\
  -a_{1}a_{5}/b & a_{1} & 0 & 0 \\
  a_{2}a_{4} & 0 & a_{2} & 0 \\
  a_{3} & a_{4} & a_{5} & 1 \\
\end{array}
\right)$  \\
& $,f^{3}_{34}=b,f^{1}_{23}=1$\\
 \hline
\vspace{-4mm}
 $A^{1}_{4,9}$ &
$f^{1}_{14}=2,f^{2}_{24}=1$ & $\left(
\begin{array}{cccc}
  a_{1}a_{4}-a_{2}a_{3} & 0 & 0 & 0 \\
  a_{2}a_{6}-a_{1}a_{7} & a_{1} & a_{2} & 0 \\
  a_{4}a_{6}-a_{3}a_{7} & a_{3} & a_{4} & 0 \\
  a_{5} & a_{6} & a_{7} & 1 \\
\end{array}
\right)$  \\
& $,f^{3}_{34}=1,f^{1}_{23}=1$\\
 \hline
\vspace{-4mm}
  $A^{0}_{4,9}$
&$f^{1}_{14}=1,f^{2}_{24}=1$ & $\left(
\begin{array}{cccc}
  a_{2}a_{3} & 0 & 0 & 0 \\
  a_{1} & a_{2} & 0 & 0 \\
 a_{3}a_{5} & 0 & a_{3} & 0 \\
  a_{4} & a_{5} & 0 & 1 \\
\end{array}
\right)$\\
&  $,f^{1}_{23}=1$\\
 \hline
 \vspace{-4mm}
 $A_{4,10}$ & $f^{3}_{24}=-1,f^{2}_{34}=1$ &
$\left(
\begin{array}{cccc}
  a^2_{1}+a^2_{2} & 0 & 0 & 0 \\
  -a_{1}a_{4}-a_{2}a_{5} & a_{1} & a_{2} & 0 \\
  a_{2}a_{4}-a_{1}a_{5} & -a_{2} & a_{1} & 0 \\
  a_{3}& a_{4} & a_{5} & 1 \\
\end{array}
\right)$\\
&  $,f^{1}_{23}=1$\\
 \hline
 \vspace{-4mm}
 $A^{a}_{4,11}$ & $f^{1}_{14}=2a,f^{2}_{24}=a,f^{3}_{24}=-1$ &  $\left(
\begin{array}{cccc}
  a^{2}_{1}+ a^{2}_{2} & 0 & 0 & 0 \\
  -\frac{a(a_{1}a_{4})+a(a_{2}a_{5})+a_{2}a_{4}-a_{1}a_{5}}{a^{2}+{1}}  & a_{2} & -a_{1}  & 0 \\
   \frac{a(a_{2}a_{4})-a(a_{1}a_{5})-a_{1}a_{4}-a_{2}a_{5}}{a^{2}+{1}} & a_{1} & a_{2} & 0 \\
  a_{3} & a_{4} & a_{5} & 1 \\
\end{array}
\right)$  \\
&  $,f^{2}_{34}=1,f^{3}_{34}=a,f^{1}_{23}=1$\\
 \hline
\vspace{-4mm}
 $A_{4,12}$&
$f^{2}_{14}=-1,f^{1}_{13}=1$ & $\left(
\begin{array}{cccc}
  a_{2} & -a_{1} & 0 & 0 \\
  a_{1} & a_{2} & 0 & 0 \\
  -a_{4} & a_{3} & 1 & 0 \\
  a_{3} & a_{4} & 0 & 1 \\
\end{array}
\right)$  \\
&  $,f^{1}_{24}=1,f^{2}_{23}=1$\\
 \hline
\end{tabular}
\end{center}
\newpage
\begin{eqnarray}
\Lambda_{1}=\textrm{Rotation}_{xy}\textrm{Boost}_{xz}\textrm{Boost}_{yz}C
\end{eqnarray}
where:
\begin{eqnarray}
\textrm{Rotation}_{xy}&=&\left(
\begin{array}{cccc}
   \cos(a_{1}) & \sin(a_{1}) & 0 & 0 \\
   -\sin(a_{1}) & \cos(a_{1}) & 0 & 0 \\
   0 & 0 & 1 & 0 \\
   0 & 0 & 0 & 1
\end{array}
\right)\\
\textrm{Boost}_{xz}&=&\left(
\begin{array}{cccc}
  \cosh(a_{2}) & 0 & \sinh(a_{2}) & 0\\
   0 & 1 & 0 & 0 \\
  \sinh(a_{2}) & 0 & \cosh(a_{2}) & 0 \\
   0 & 0 & 0 & 1
\end{array}
\right)\\
\textrm{Boost}_{yz}&=&\left(
\begin{array}{cccc}
   1 & 0 & 0 & 0\\
   0 & \cosh(a_{3}) & \sinh(a_{3}) & 0 \\
   0 & \sinh(a_{3}) & \cosh(a_{3}) & 0 \\
   0 & 0 & 0 & 1
\end{array}
\right)\\
C&=&\left(
\begin{array}{cccc}
  1 & 0 & 0 & 0\\
  0 & 1 & 0 & 0 \\
  0 & 0 & 1 & 0 \\
  0 & 0 & 0 & a_{4}
\end{array}
\right)\\
\end{eqnarray}
\begin{eqnarray}
\Lambda_{2}=\textrm{Rotation}_{xy}\textrm{Rotation}_{xz}\textrm{Rotation}_{yz}C
\end{eqnarray}
where:
\begin{eqnarray}
\textrm{Rotation}_{xy}&=&\left(
\begin{array}{cccc}
   \cos(a_{1}) & \sin(a_{1}) & 0 & 0 \\
   -\sin(a_{1}) & \cos(a_{1}) & 0 & 0 \\
   0 & 0 & 1 & 0 \\
   0 & 0 & 0 & 1
\end{array}
\right)\\
\textrm{Rotation}_{xz}&=&\left(
\begin{array}{cccc}
  \cos(a_{2}) & 0 & -\sin(a_{2}) & 0\\
   0 & 1 & 0 & 0 \\
  \sin(a_{2}) & 0 & \cos(a_{2}) & 0 \\
   0 & 0 & 0 & 1
\end{array}
\right)\\
\textrm{Rotation}_{yz}&=&\left(
\begin{array}{cccc}
   1 & 0 & 0 & 0\\
   0 & \cos(a_{3}) & \sin(a_{3}) & 0 \\
   0 & -\sin(a_{3}) & \cos(a_{3}) & 0 \\
   0 & 0 & 0 & 1
\end{array}
\right)\\
C&=&\left(
\begin{array}{cccc}
  1 & 0 & 0 & 0\\
  0 & 1 & 0 & 0 \\
  0 & 0 & 1 & 0 \\
  0 & 0 & 0 & a_{4}
\end{array}
\right)
\end{eqnarray}

\end{document}